\documentclass[aps,prd,reprint,twocolumn,preprintnumbers,floatfix,nofootinbib]{revtex4-1}
\usepackage{graphicx}
\usepackage{amsmath}
\usepackage{amsfonts}
\usepackage{ulem}
\usepackage{multirow}
\usepackage{placeins}
\usepackage{mathtools}
\usepackage{todonotes}
\usepackage{physics}
\usepackage{amssymb}
\usepackage{bm}
\usepackage{csquotes}
\usepackage{epstopdf}
\usepackage{float}
\usepackage{xcolor}

\usepackage{capt-of}
\allowdisplaybreaks
\usepackage{natbib}

\usepackage{hyperref}

\hypersetup{
     colorlinks   = true,
     citecolor    = red,
     linkcolor    = blue,
     urlcolor     = blue,
}
\definecolor{deeppink}{rgb}{0.9, 0.17, 0.31}

\def\beq{\begin{equation}}
\def\eeq{\end{equation}}
\def\bea{\begin{align}}
\def\eea{\end{align}}

\def\l{\left}
\def\r{\right}

\def\ae{a_{\rm e}}

\def\ke{k_{\rm e}}
\def\He{H_{\rm e}}

\def\HI{H_{\rm e}}
\def\Hre{H_{\rm re}}
\def\Tre{T_{\rm re}}

\def\MP{{ M}_{\rm pl}}

\def\Tre{T_{\rm re}}

\def\mJ_{\mathrm{J}}
\def\nno{\nonumber}

\graphicspath{{./figs/}}
\begin{document}

 \title{
 %WIMP and FIMP with the long-wavelength DM fluctuations
% \\Infrared contribution to WIMP and FIMP Dark Matter \\
%WIMPy and FIMPy Infrared Dark Matter\\
%Dark matters are completely dark or WIMPy or FIMPy, or UFOy: An Inflation Dominion\\
Dark matters are Inert, or FIMPy, or WIMPy or UFOy: \\An inflationary gravitational particle production}

%\iffalse
 \author{Ayan Chakraborty}
\email{E-mail:chakrabo@iitg.ac.in}
\affiliation{Department of Physics, Indian Institute of Technology, Guwahati, 
Assam, India}
\author{Debaprasad Maity}
\email{E-mail: debu@iitg.ac.in}
\affiliation{Department of Physics, Indian Institute of Technology, Guwahati, 
Assam, India}
 \author{Rajesh Mondal}
\email{E-mail: rajeshm@iiserbpr.ac.in}
\affiliation{Department of Physics, Indian Institute of Technology Guwahati, Assam, India}
\affiliation{Department of Physical Science, Indian Institute of Science Education and Research Berhampur, Berhampur 760003, Odisha, India}

 \begin{abstract}
 In this letter, we explore the phenomenological impact of inflationary gravitational particle production in the physics of Dark Matter (DM).
 %an inescapable natural portal for the production of thermal and non-thermal scalar Dark Matter (DM) during the universe's evolution from the early inflation to the post-inflationary radiation-dominated era 
 %In the framework of Cosmological Gravitational Particle Production (CGPP). 
 %The existence of a shrinking comoving Hubble horizon during inflation 
Large-scale DM fluctuations generated during inflation behave as gravitational particles upon their post-inflationary horizon reentry and alter the conventional Boltzmann dynamics of DM  with a non-conserving source term, thereby producing significant phenomenological consequences.
%Those gravitationally produced particles are shown to 
%a significant phenomenological impact. 
%in a significant way.
%alter the thermal (WIMP, UFO) and non-thermal (FIMP) DM phenomenology. 
%We consider the gravitationally produced large-scale massive scalar number density spectrum during the inflation to radiation transition. 
%The interaction between the large scales of DM field and the post-inflationary thermal bath is captured by introducing a production rate($Q_{\rm IR}$) in addition to the thermal bath production rate in the Boltzmann equation governing the evolution of DM number density. This $Q_{\rm IR}$ accounts for the phenomenon of successive Hubble horizon entry of IR modes of the DM field, and their contribution to the DM number density evolution during the post-inflationary era. 
Within this framework, we analyze four distinct types of DM classified according to their production mechanisms. Dark matter may be completely non-interacting with the thermal bath, behaving as Inert Dark Matter. Alternatively, depending on the strength of its interactions with bath particles, DM may exhibit WIMPy, UFOy, or FIMPy behavior, sharing characteristics with their conventional counterparts. The late-time enhancement of the DM number density, driven by the successive horizon reentry of gravitationally produced low-momentum modes, enlarges the viable parameter space for both thermal and non-thermal DM scenarios. Remarkably, this expanded parameter space remains consistent with current constraints from $\Delta N_{\rm eff}$ and Lyman-$\alpha$ bound. %It extends the mass of thermal DM into MeV regime , which is in concordance with the $\Delta N_{\rm eff}$ constraint at the time of Big Bang nucleosynthesis(BBN) and the Lyman-$\alpha$ warmness constraint at the time of structure formation around $T\sim 1$ eV. 
 %The present system within the CGPP framework dictates the viability of the UFO mechanism even in the instantaneous reheating scenario.            
 \end{abstract}
% \fi
\maketitle

\section{\textbf{ Introduction }}\label{secintro}
Dark matter is one of the greatest puzzles in modern cosmology. Its veiled nature is so profound that nearly seven decades of persistent indirect observational evidence have not led to any major scientific breakthrough. Meanwhile, several decades of direct detection efforts have only rendered its existence more elusive. Despite numerous cosmological experiments with ever-increasing precision, we have obtained only a few definitive pieces of information: the current dark matter abundance is approximately $\Omega_{\chi} h^2 \simeq 0.12$, it must be massive, and a significant portion of it must be cold in nature. The observed structure formation, from Galaxies to Galaxy clusters, cannot be explained within the framework of the standard cosmological model($\Lambda$CDM) without DM. Moreover, precise measurements of the Cosmic Microwave Background(CMB) \cite{Planck:2018jri}, gravitational lensing \cite{Clowe_2006, Yoo_2010, Mellier:1998qu}, and the dynamics
of galaxies and clusters \cite{SDSS:2003eyi} all independently point toward the presence of a non-luminous and non-baryonic matter component. In the present decade, the search for the nature of DM has become one of the central
questions at the nexus of cosmology, astrophysics, and high-energy particle physics.

Over the years, several theoretical frameworks have been proposed to uncover the true nature of dark matter. It could be a beyond-the-Standard-Model (BSM) particle with weakly/feebly interacting with Standard Model counterparts \cite{Arcadi:2017kky,Bernal:2017kxu}; it might manifest as an effective background geometry within a modified gravity framework \cite{Capozziello:2006dj}; or it could be a purely classical field such as the axion \cite{Preskill:1982cy,Abbott:1982af,Dine:1982ah}.

In this letter, we remain within the realm of particle physics and pose the following question: even within the simplest DM model frameworks, have all possible channels of dark matter production been fully explored? Recent developments in the study of gravitational dark matter suggest that the answer is no. One important production mechanism that has been largely overlooked until recently involves dark matter generation through universal gravity-mediated scattering processes, particularly via the interaction term $h_{\mu\nu} T^{\mu\nu}$. However, it has been realized that inflation plays a crucial role in making this production channel phenomenologically significant. Due to its high energy content, the inflaton field can copiously produce dark matter particles during and after inflation through gravitational interactions. A detailed analysis of this novel DM production mechanism and its phenomenological implications has been carried out recently in a series of papers by various authors\cite{Barman:2022qgt, Haque:2022kez, Haque:2023yra, Barman:2023opy, Bernal:2018qlk,Mambrini:2021zpp,Clery:2021bwz, Ahmed:2020fhc,Kaneta:2023uwi,Choi:2024ilx, Garcia:2020eof,Moroi:2020bkq,Haque:2022kez,Haque:2021mab,Clery:2021bwz, Haque:2023zhb}. The discovery of such a channel reveals the fact that in any dark matter phenomenology, such a gravitational contribution happens to be inescapable.\\
%should always be taken into account.
In this paper, we yet consider another novel gravity-mediated, non-perturbative production channel that has been explored recently \cite{Chakraborty:2025zgx, Chakraborty:2025oyj, Dolgov:1989us,Ema:2016hlw,Ema:2018ucl,Chung:2018ayg,Basso:2021whd,Hashiba:2018iff,Hashiba:2018tbu,Herring:2019hbe,Lankinen:2019ifa, Markkanen:2018gcw, Markkanen:2018gcw,Choi:2024bdn, Garcia:2025rut, Enqvist:2014zqa, Tenkanen:2019aij, Choi:2024bdn}, but with its potential implications for DM phenomenology still being uncovered. The framework is the well-known gravitational particle production in a time-dependent background. Inflation plays a key role in this regard. The inflationary epoch is well known to be an ideal laboratory for gravitational particle production. One of its most profound consequences is the late-time structure formation, which can be traced back to the infrared fluctuations of the inflaton field, interpretable as very low-energy quantum inflaton particles produced by the inflationary background. Due to its very gravitational nature, such production of infrared fluctuation applies to any quantum field, such as DM, which is our present topic of discussion. In this letter, we indeed demonstrate that such inescapable and universal gravitational production profoundly alters the DM phenomenology.

%This paper is constructed as follows: In Section \ref{secDMspectrum}, we briefly review the non-perturbative inflationary gravitational particle production, and compute the large-scale number density spectrum of the massive scalar DM field during the inflation to radiation transition in the early universe. In Section \ref{secDMpheno}, we illustrate the thermal and non-thermal DM phenomenology, exploring four distinct possibilities of DM being non-interacting(inert), or FIMPy, or WIMPy, or UFOy. We conclude this section by presenting a rigorous discussion on the key features of the aforesaid mechanisms in the present system, and also discuss the relevant observational bounds, namely $\Delta N_{\rm eff}$ and Lyman-$\alpha$ bounds, which finally constrain the WIMP, FIMP, and UFO parameter space.  

\section{Inflationary Gravitational particle production }\label{secDMspectrum}
%Let us begin our discussion with the basic mechanism of universal gravitational particle production. 
To prepare the stage, we consider the simplest scenario with inflation followed by the radiation phase with instantaneous reheating. Inclusion of a non-trivial reheating phase will be studied later. We consider the following non-minimally coupled massive($m_{\chi}$) scalar dark matter field($\chi$) Lagrangian as, %%%%%%%%%%%%%%%%%%%%%%%%%%%%
\begin{equation}\label{lagrangian1}
    \mathcal{L}_{\chi}=-\sqrt{-g}\l(\frac{1}{2}\partial_{\mu}\chi \partial^{\mu}\chi
    +\frac{1}{2}(m_{\chi}^2 + \xi R)\chi^2\r).
  \end{equation}
The background FLRW metric is expressed as $ds^2=a^2(\eta)\big(-d\eta^2+d\vec{x}^2\big)$\footnote{ (-,+,+,+) metric signature is followed throughout.} with $\sqrt{-g}=a^4(\eta)$. 
The rescaled Fourier mode of the scalar field $a(\eta)\chi_k$ satisfies the following dynamical equation,
%%%%%%%%%%%%%%%%%%%%%%%%%%
   \begin{equation}\label{dynamical2}       X_{{k}}^{\prime\prime}+ \omega^2_k(\eta)X_{{k}}=0~~;~~ \omega^2_k(\eta) =k^2+a^2 m_{\chi}^2 -\frac{a^{\prime\prime}}{a}(1+6\xi) .
  % =  \bigg(k^2+a^2m_{\chi}^2-\frac{a^{\prime\prime}}{a}\bigg)
  \end{equation}
%%%%%%%%%%%%%%%%%%%%%%%%%%
Note that while an inflationary background admits an adiabatic Bunch-Davies(BD) vacuum in the remote past $\eta \rightarrow -\infty$, the post-inflationary radiation-dominated(RD) universe also admits an adiabatic vacuum in the distant future $\eta \rightarrow \infty$. It is the intermediate nontrivial time-dependent background that leads to the mixing of positive
and negative frequency eigenmodes. When all the modes are deep inside the horizon, one can decompose $X_k(\eta)$ as
%%%%%%%%%%%%%%%%%%%%%%%%%%%%%%
\begin{eqnarray}
X_k\simeq \left\{\begin{array}{ll}
\frac{1}{\sqrt{2 \omega^{\rm (inf)}_k}}\left(
a_{\vec k} e^{-i\int \omega_k d\eta'}  + 
a_{-\vec k}^\dagger e^{i\int \omega_k d\eta'}\right) & \eta \rightarrow -\infty\\
\frac{1}{\sqrt{2 \omega^{\rm (rad)}_k}}\left(
b_{\vec k} e^{-i\int \omega_k d\eta'}  + 
b_{-\vec k}^\dagger e^{i\int \omega_k d\eta'}\right) & \eta \rightarrow \infty.
\end{array}
\right.
\label{Eq:Xx}
\end{eqnarray}
%%%%%%%%%%%%%%%%%%%%%%%%%%%%%%%
%with 
%\begin{equation}
%\Omega_k(\eta)=\int^\eta\omega_k(\eta')d\eta'\,,
%\end{equation}
Where $(a_{\vec k},a_{\vec k}^{\dagger})$ and $(b_{\vec k}, b_{\vec k}^{\dagger)})$ are the creation and annihilation operators associated with two independent adiabatic vacua, defined in inflation and radiation background, respectively. 
%$\omega^{(\rm  inf)/(\rm  rad)}_k$ are time-dependent frequencies during inflation and radiation-dominated phase, respectively. 
The  Bunch-Davies vacuum is defined by $a_k|0\rangle_{\rm BD}=0$, and $b_k|0\rangle_{\rm BD}=0$.  
The standard Bogoliubov approach states that those two distinct vacua can be related via a unitary transformation as follows,
%%%%%%%%%%%%%%%%%%%%%%%%
\beq
b_{\vec k}=\alpha_{\vec k} a_{\vec k}+\beta^*_{\vec k} a^\dagger_{-\vec k}\,,
~~ 
b^\dagger_{-\vec k}=\alpha^*_{\vec k}a^\dagger_{-\vec k}+\beta_{\vec k}a_{\vec k}\,.
\eeq
%%%%%%%%%%%%%%%%%%%%%%%%%
Where $\alpha_{\vec{k}},~ \beta_{\vec{k}}$ are the 
(time-dependent) Bogoliubov coefficients satisfying the normalization condition\footnote{Extracted from the Wronskian condition on $X(\eta,x)$, 
$\l(X_k\,X_k^{*'}-X_k^* \,X_k'\r)=i$.} $|\alpha_{\vec{k}}|^2-|\beta_{\vec{k}}|^2=1$. With this new set of operators, one deduces 
number density of the produced particles as
\beq \label{nDM}
 n_\chi=\frac{1}{a^3}
\int\frac{d^3k}{(2 \pi)^3}
 {_{\rm BD}\langle 0 | b_{-\vec k}^\dagger b_{\vec k}|0\rangle_{\rm BD}}=
\int \frac{d^3k}{(2 \pi)^3}
|\beta_k|^2\,,
\eeq
%%%%%%%%%%%%%%%%%%%%%%%%%%%%%%%%%%%%%%
where we defined the initial state at time $\eta \rightarrow -\infty$ by $a_{\vec k}|0\rangle=0$, or
$\beta_{\vec k}| \rightarrow 0$.
Note also that the occupation number, $|\beta_k|^2$, is equivalent
to a distribution function $f_\chi(|k|,t)$ in the Boltzmann
approach \cite{Garcia:2022vwm, Chakraborty:2025zgx, Chakraborty:2024rgl, Chakraborty:2025oyj, Kaneta:2022gug}. 

For our present study, we parametrize the form of the massive scalar dark matter particle spectrum in the long-wavelength limit as follows\cite{Chakraborty:2025zgx, Herring:2019hbe}
%%%%%%%%%%%%%%%%%%%%%%%%%%%%%
%\begin{subequations}\label{exactbeta}
\begin{align}\label{exactbeta}
 % |\beta_k|_{\rm exact}^2\approx& \frac{e^{-\frac{\pi \He }{4 m_{\chi}}\left(\frac{k}{\ke}\right)^2}}{\sqrt{m_{\chi}/\He}}\left(\frac{k}{\ke}\right)^{\delta}\,\\
   |\beta_k|^2\approx \frac{{\cal{A}}\,e^{-\frac{\pi \He }{4 m_{\chi}}\left(\frac{k}{\ke}\right)^2}}{\sqrt{m_{\chi}/\He}}\left(\frac{k}{\ke}\right)^{\delta}\,~\text{for}~~ k^2 \He < \ke^2 m_{\chi},
\end{align}
%\end{subequations}
%%%%%%%%%%%%%%%%%%%%%%%%%%%%
which is assumed to be excited as the DM field evolves from inflation to radiation dominated phase. For example, DM with non-minimal coupling $\xi R \chi^2$, the spectral index sub-Hubble mass limit assumes the following approximate form  \cite{Chakraborty:2025zgx} %\l(m^2_{\chi}/\He^2+12\xi\r)<9/4$.
 $\delta\sim\,-\sqrt{9-48\xi-{4m_{\chi}^2}/{\He^2}}$, where $|\delta_{\rm max}|=3$, $\He$ is the Hubble scale at the end of inflation, and $\ke=\ae\He$ is the scale that left the horizon at the inflation end with $\ae$ being the associated scale factor. The amplitude $\mathcal{A}$ will be approximately of the order of unity in the range $ k^2 \He < \ke^2 m_{\chi}$. 
%For If we choose $m_{\chi}<<\He$, $g< 10^{-10}$, or $\alpha<10^{-10}\MP$ considering $\phi_0\approx \MP$, and $\He\simeq 10^{-5} \MP$, three terms in the inflationary effective mass $m^{\rm inf}_{\rm eff}$ become ineffective except the non-minimal coupling assisted term. 
%Now the index $\delta$ can be greater than -3, if $\xi$ is negative. 
%For our present purpose we choose the value $\delta \geq -3$.
%So, the only possibility is the negative non-minimal coupling $\xi<0$. It is strong inflationary instability caused by the negative $\xi$, which causes heavier IR tilt of the spectrum \cite{Chakraborty:2024rgl}. This is the practical situation of obtaining the IR spectral index $|\delta|>3$ by varying the negative non-minimal coupling strength.\\
However, to maintain generality, we consider $\delta$ as a free parameter, and we shall restrict ourselves within the range $|\delta|\leq 3$ for $|\delta|> 3$, the energy density of DM field, $m_{\chi}\,k^3|\beta_k|^2$, becomes IR divergent, causing the early matter domination long before the radiation-matter equality. In this regard, we would like to mention that such an IR-convergent DM energy density spectrum in the range $|\delta|\lesssim 3$ is also consistent with the current CMB scale isocurvature bound \cite{ACT:2025tim, ACT:2025fju}, as extensively discussed in \cite{Chakraborty:2024rgl, Chakraborty:2025oyj}.  

We now attempt to propose a quantitative description of the Infrared (IR) production rate of massive dark matter fluctuations. As stated before, after their inflationary horizon exit, all the modes $k < \ke$ (infrared modes) reenter the Hubble horizon during the post-inflationary radiation period, and contribute to the DM particle number density. This fact motivates us to define an equivalent dark matter production rate associated with these IR modes as a function of growing Hubble horizon size, capturing their gradual entry throughout the subsequent phases of evolution. 
Our main objective would be to explore DM phenomenology, taking into account this universal IR contribution. 
%behind this computation is to capture the effect of the time-varying number density of dark matter particles due to gradual long-wavelength mode entry throughout the subsequent phases of evolution.
Depending on its mass and  coupling with the standard model particles, we discuss four possibilities: DM can be completely dark(gravitational), which we call Inert, 
%non-relativistic,
Feebly Interacting Massive Particle-like (FIMPy), Weakly Interacting Massive Particle-like (WIMPy), and intermediate Ultra-relativistic Freeze-out-like (UFOy). 
%Recently realised UFO-like possibility arises in the low relativistic-mass regime, where DM particles get decoupled from the standard model thermal bath relativistically when the bath temperature remains high enough compared to the DM mass. %This is known as ultra-relativistic freeze-out(UFO). 
%In this letter we explore all these distinct possibilities.   
%scenario in a pure radiation-dominated(RD) universe after the end of the early accelerated inflationary era. 
%Therefore, we shall study the time-varying number density of DM particles caused by the IR contribution during the radiation-dominated phase until the largest scale(here we consider it to be $k_{\rm CMB}$) enters the horizon at some point near the radiation-matter equality.

\section{Dark Matter Phenomenology}\label{secDMpheno}

\subsection{DM non-interacting with the thermal bath}
{\bf Computation of IR production rate \enquote{$Q_{\rm IR}$}:} Without any standard model and inflaton field interaction, the appropriate  Boltzmann equation governing the evolution of dark matter number density($n_{\chi}$) can straightforwardly follow from the equation (\ref{nDM}),
%%%%%%%%%%%%%%%%%%%%%%%%%%%
\begin{align}\label{eq:nchievolution1}
\frac{1}{a^3}\frac{d N_\chi}{dt} = Q_{\rm IR} = %-\langle\sigma v\rangle\left(n_{\chi}^2-n_{\rm eq}^2\right)
  \frac{1}{a^3}\frac{d}{dt} \int_{a(t) H(t)}^{\ke}\frac{d^3 k}{(2\pi)^3}|\beta_k|^2\,,
\end{align}
%%%%%%%%%%%%%%%%%%%%%%%%%%%%%%
where, \enquote{dot} is defined with respect to cosmic time $dt = a(\eta)d\eta$. Total DM number $N_\chi = a^3 n_{\chi}$. Note the lower limit ($k = a(t)H(t)$) in the right-hand side integral indicating the fact that at any instant of time during the post-inflationary evolution, the modes from $\ke \to a(t)H(t)$ will enter the horizon. Those are the modes which will contribute to the evolution of the DM particle number density, and  
%However, this picture can be imprinted on the dark matter number density once we solve the Boltzmann equation by modifying its right-hand side with an additional term,
we call it as \textit{Infrared production rate}, $Q_{\rm IR}$. 
Utilizing the generic spectrum (\ref{exactbeta}), the expression of the IR production rate 
%based on the generic spectrum (\ref{exactbeta}) 
is computed as, 
%in terms of the normalized scale factor $\bar{a}\equiv(a/\ae)$ as follows:
%%%%%%%%%%%%%%%%%%%%%%%%%
\begin{equation}\label{eq:genericQIR}
\boxed{Q_{\rm IR}= \frac{\mathcal{-A}\He^{\frac52}}{2 \pi^2 {m_{\chi}^{\frac12}}}\frac{H}{\bar{a}^2}\l(\frac{\bar{a} H}{\He}\r)^{2+\delta}e^{-\frac{\pi \He }{4 m_{\chi}}\left(\frac{\bar{a}\,H}{\He}\right)^2} \frac{d}{d\bar{a}}\l(\bar{a} H\r)}\,.
\end{equation}
%%%%%%%%%%%%%%%%%%%%%%%
Where, the normalized scale factor $\bar{a}\equiv(a/\ae)$. 
We now attempt to explore the impact of this infrared production, and demonstrate its importance on DM phenomenology for a large range of parameter space particularly for $\delta < 0$. %The heavier the red tilt, the more the importance of $Q_{\rm IR}$. 
This is the most important expression of our analysis. 

In this work, we primarily focus on the instantaneous reheating case. Inflation is therefore followed by the radiation-domination with the Hubble scale $H$ behaves as, $H(\bar{a})=\Hre\l(\bar{a}/\bar{a}_{\rm re}\r)^{-2}$. 
For instantaneous reheating scenario, the Hubble scale at the reheating end is $\Hre=\He=\l(\Tre^2/\MP\r)\sqrt{\frac{\epsilon}{3}}$, with $\epsilon=\l(\pi^2 g_{\ast}/30\r)$, where $g_{\ast}$ is the number of relativistic degrees of freedom contributing to the Standard Model(SM) energy density in the thermal bath, and we have taken it $427/4\approx 106.75$. For standard large-scale inflation, the instantaneous transition from inflation to the RD phase predicts the reheating temperature $\Tre \approx 10^{15}$ GeV and the normalized scale factor at the reheating end, $\bar{a}_{\rm re}\approx 1$. We will use these values in the subsequent analysis.
%%%%%%%%%%%%%%%%%%%%%%%%%%%%%%
\begin{figure}
\begin{center}
\includegraphics[scale=0.255]{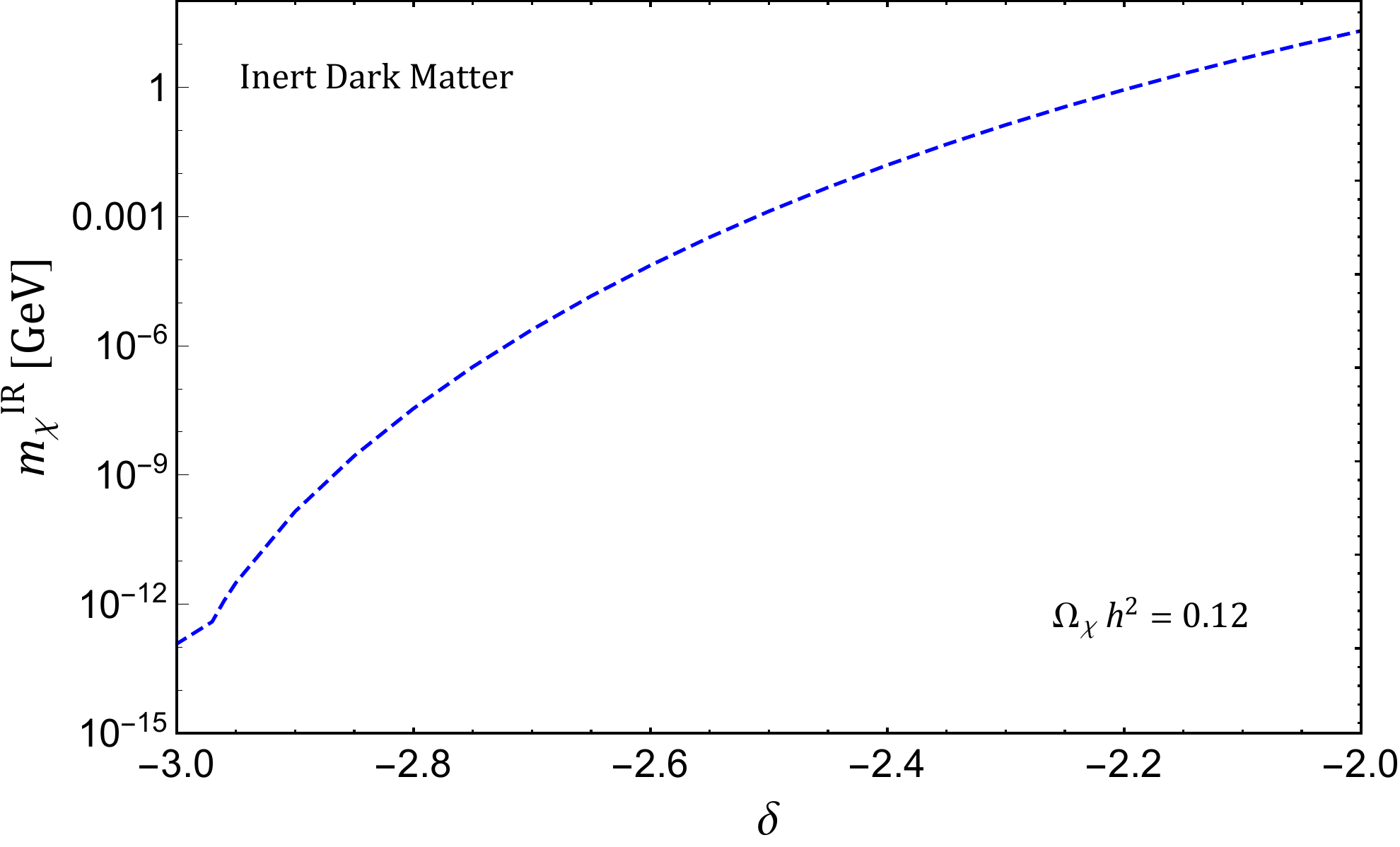}
\caption{\textit{Figure represents IR mass $m_{\chi}^{\rm IR}$ vs $\delta$ parameter space for purely gravitational dark matter or an inert dark matter. Each $\delta$ gives a single DM mass which satisfies the present relic.}}
\label{figmvsdeltapuregrav}
\end{center}
\end{figure}
%%%%%%%%%%%%%%%%%%%%%%%%%

\textbf{Inert Dark Matter:} Subject to these parameters during the radiation-dominated phase, Eq.(\ref{eq:genericQIR}) boils down to the following form,
%%%%%%%%%%%%%%%%%%%%%%%%%%%%%%
\begin{equation}\label{eq:genericQIRRD}
   Q_{\rm IR}= \frac{\mathcal{A}\,\He^{9/2}}{2 \pi^2 \sqrt{m_{\chi}}}\frac{e^{-\frac{\pi \He }{4 m_{\chi}\bar{a}^2}}}{\bar{a}^{(8+\delta)}}\,.
\end{equation}
%%%%%%%%%%%%%%%%%%%%%%%%%%%%%%%%%%%%
%%%%%%%%%%%%%%%%%%%%%%%%%%%%%%%%%%%
%\textcolor{red}{Give analytic expression of DM abundance, and some estimate of DM mass here} 
%With the development of this long-wavelength DM production rate, we now proceed to study the well-known freeze-in and freeze-out scenarios of dark matter.
Using Eq.(\ref{eq:genericQIRRD}) into Eq.(\ref{eq:nchievolution1}) one performs  the integration and obtains total IR contribution to the dark matter as,
%%%%%%%%%%%%%%%%%%%%%%%%%%%%%%%
\begin{align}\label{eq:nchisolIRFI}
   N^{\rm IR}_{\chi}
   %& \frac{\mathcal{A}\He^{7/2}}{2\pi^2\sqrt{m_{\chi}}}\int_{1}^{\bar{a}_{\rm IR}}\frac{e^{-\frac{\pi\He}{4m_{\chi}\bar{a}^2}}}{\bar{a}^{(4+\delta)}}d\bar{a}\nno\\
%   &=\frac{\mathcal{A}\He^{7/2}}{4\pi^2\sqrt{m_{\chi}}}\l(\frac{\pi\He}{4m_{\chi}}\r)^{-\frac{(3+\delta)}{2}}\nno\\
%   &\l(\Gamma\l[\frac{3+\delta}{2},\,\frac{\pi\He}{4m_{\chi}\bar{a}_{\rm IR}^{2}}\r]-\Gamma\l[\frac{3+\delta}{2},\,\frac{\pi\He}{4m_{\chi}}\r]\r)\\
 %  &=\mathbf{\frac{\mathcal{A}\He^{7/2}}{4\pi^2\sqrt{m_{\chi}}}\l(\frac{\pi\He}{4m_{\chi}}\r)^{-\frac{(3+\delta)}{2}}\,\Gamma\l[\frac{3+\delta}{2},\,\frac{\pi\He}{4m_{\chi}\bar{a}_{\rm IR}^{2}}\r]}\\
% \simeq& \frac{\mathcal{A}\HI^{7/2}}{4\pi^2\sqrt{m_\chi}}\left(\frac{4m_\chi}{\pi\HI}\right)^{\frac{3+\delta}{2}}\Gamma\left[\frac{3+\delta}{2},\,\frac{\pi\He}{4m_{\chi}\bar{a}_{\rm IR}^2}\right]\nno\\
\simeq \frac{\mathcal{A}\He^{7/2}}{4\pi^2\sqrt{m_{\chi}}}
%&\times
\begin{cases}
{\rm ln}\l(\frac{4\, m_{\chi}\,\bar{a}^2_{\rm IR}}{\pi\,\He}\r)~~~~\quad\quad\quad\mbox{for}~~\delta=-3\\
\l(\frac{4m_{\chi}}{\pi\He}\r)^{\frac{(3+\delta)}{2}}\,\Gamma\l[\frac{3+\delta}{2}\r]~~\mbox{for}~~|\delta|<3\,,
\end{cases}
\end{align}
%%%%%%%%%%%%%%%%%%%%%%%%%%%%%
%where $\Gamma[\alpha,\,\beta]$ is the incomplete gamma function involving two arguments $\alpha$ and $\beta$, $\Gamma[\alpha]$ is the gamma function with one argument $\alpha$, and 
where $\bar{a}_{\rm IR}$ is the normalized scale factor at the present time. For our present purpose, we chose it to be around the instant of matter-radiation equality  
%is some late time during the RD phase when both radiation and dark matter energy density freeze. 
$\bar{a}_{\rm IR}\approx 2.44\times 10^{26}\l(\frac{\He}{\Tre}\r)$. %is chosen to be the instant around radiation-matter equality, 
%\textcolor{red}{Need to understand the following point::: when the CMB pivot scale, $k_{\ast}/a_0=0.05\,\text{Mpc}^{-1}$, enters the horizon.} 
%Due to universal inflationary IR contribution, we, therefore, have the total comoving DM number density, $N_{\chi}\approx \l(N^{\rm UV}_{\chi}+N^{\rm IR}_{\chi}\r)$.\\

The present-day relic abundance of the Inert DM can therefore be straightforwardly written as 
%%%%%%%%%%%%%%%%%%%%%%%%%%%
\begin{equation}\label{eq:relicpuregrav}
%\begin{aligned}
\Omega^{\rm Inert}_\chi h^2= \frac {m_{\chi} N_{\chi}^{\rm IR}}{\epsilon\,\Tre^3\,T_0} \Omega_{\rm R}h^2 ,
%\nno\\
%&\left( \textcolor{black}{\frac{\mathcal{A}\HI^{7/2}\sqrt{m_{\chi}}}{4\pi^2\epsilon\,\Tre^3\,T_0}\left(\frac{4m_\chi}{\pi\HI}\right)^{\frac{3+\delta}{2}}\Gamma\left[\frac{3+\delta}{2},\,\frac{\pi\He}{4m_{\chi}\bar{a}_{\rm IR}^2}\right]}\right)\Omega_{\rm R}h^2\,,
%\end{aligned}
\end{equation}
%%%%%%%%%%%%%%%%%%%%%%%%%%
which is parametrized by the present-day radiation energy density parameter, $\Omega_{\rm R}h^2=4.3\times 10^{-5}$ \cite{Planck:2018jri, Planck:2018vyg}, and the present-day CMB temperature, $T_0\simeq2.35\times 10^{-13}$ GeV, defined at the present-day scale factor $a_0$. From this expression, we determine the critical IR mass($m_{\chi}^{\rm IR}$) satisfying the current relic, $\Omega_{\chi}h^2\simeq 0.12$ \cite{Planck:2018vyg, ParticleDataGroup:2024cfk, Bernal_2022}, as follows:
%%%%%%%%%%%%%%%%%%%%%%%%%%
\begin{align}\label{eq:mIR}
&m^{\rm IR}_\chi\simeq
\begin{cases}
(2\times 10^{-13}\,{\rm GeV}^2)\frac{\Tre^6}{\mathcal{A}^2\,\He^7\,\mathcal{W}_0(q_2)}~~~~%\quad\quad\quad\quad\quad~~~~~~~~~
\mbox{for}~~\delta=-3\\
\left(\frac{(9\times10^{-7}{\rm GeV})\,\Tre^3}{ \mathcal{A}\,\Gamma\left[\frac{3+\delta}{2}\right]\HI^{7/2}}\right)^{\frac{2}{4+\delta}}\,\left(\frac{\pi\,\HI}{4}\right)^{\frac{3+\delta}{4+\delta}} %\l(\mathcal{A}\,\Gamma\Big{[}\frac{3+\delta}{2}\Big{]}\r)^{-\frac{2}{4+\delta}}~
~~\mbox{for}~~|\delta|<3\,,  
\end{cases}
\end{align}
%%%%%%%%%%%%%%%%%%%%%%%%%%%%%%%%%%
where $\mathcal{W}_0(q_2)$ is the Lambert function of branch 0 with argument $q_2=(1.46\times 10^{-8}\,{\rm GeV})\frac{\Tre^3\,\bar{a}_{\rm IR}}{\He^4}$. 
%This IR mass scale also defines the characteristic mass scale for gravitationally produced pure dark matter which is not at all interacting with the thermal bath. 
%Therefore, for the dark matter being completely dark, we get a two-dimensional $m_{\chi}$ vs $\delta$ parameter space. 
For a particular spectral index $\delta$, there exists a single mass that gives the present relic, as shown in Fig.(\ref{figmvsdeltapuregrav}). For instance, we obtain the critical Inert DM mass $m_{\chi}^{\rm IR}\approx (1.18\times 10^{-13},\, 1.55\times 10^{-10},\,2.81\times 10^{-9})$ GeV for $\delta=(-3, -2.9, -2.85)$, respectively. These critical masses are much lower than the lowest mass obtained in the standard freeze-in production through the thermal bath.

\subsection{DM interacting with the thermal bath}

In the post inflationary radiation phase, as the Hubble horizon grows, the inflationary infrared DM modes gradually enter the horizon and start to interact with the thermal bath.  %contribute to the evolution of the DM number density. 
On the other hand at any particular instant of time, the modes that are outside the horizon will remain non-interacting. This fact can indeed be captured with the following IR modified Boltzmann equation for the DM number density evolution (\ref{eq:genericQIR}),
%%%%%%%%%%%%%%%%%%%%%%%%%%%
\begin{align}\label{eq:nchievolution2}
\frac{1}{\bar{a}^3}\frac{d N_\chi}{dt}  = -\frac{\langle\sigma v\rangle}{\bar{a}^6}\left(N_{\chi}^2-{N_{\chi}^{\rm eq}}^2\right)
  + Q_{\rm IR}.
\end{align} 
%%%%%%%%%%%%%%%%%%%%%%%%%%
Where $\langle\sigma v\rangle$ is the thermally averaged cross-section times velocity defining the interaction strength of dark matter particles with the thermal bath. 
%\textcolor{red}{To have an unified description let us rewrite above equation in terms of new variable $x\equiv \l(m_{\chi}/T\r)$,as, 
%%%%%%%%%%%%%%%%%%%%%%%%%%%%
%\begin{align}\label{eq:nchievolution4}
%   \frac{d}{dx}(Y_{\chi})=&-\frac{\langle\sigma v\rangle \,s}{H\,x}\l(Y_{\chi}^2-Y^{\rm eq~2}_{\chi}\r)+\frac{Q_{\rm IR}(x,\,T)}{H\,s\,x}
%\end{align}
%%%%%%%%%%%%%%%%%%%%%%%%%%%%%%
%where the DM yield $Y_{\chi}\equiv (n_{\chi}/s)$, and the SM entropy density $s(T)=\frac{2\pi^2}{45}g_{\ast s}T^3$, with $g_{\ast s}(T)$ is the number of relativistic degrees of freedom that contribute to the SM entropy.}
%The comoving DM number density evolution equation can be written in terms of \enquote{$\bar{a}$} as follows:
%%%%%%%%%%%%%%%%%%%%%%%%%%%%%%
%\begin{align}\label{eq:nchievolution3}
%  \frac{d}{d\bar{a}}(N_{\chi})=&-\frac{\langle\sigma v\rangle}{\bar{a}^4H}\l(N_{\chi}^2-N^{\rm eq~ 2}_{\chi}\r)+\frac{\bar{a}^2}{H}Q_{\rm IR}  
%\end{align}
%%%%%%%%%%%%%%%%%%%%%%%%%%%%%
%Model dependent analysis will be discussed elsewhere.
The temperature-dependent equilibrium number density $n^{\rm eq}_{\chi}\equiv (N^{\rm eq}_{\chi}/\bar{a}^3)$ for dark matter particles is known to be\cite{Bernal_2022, Haque:2023yra}, 
%%%%%%%%%%%%%%%%%%%%%%%%%%%%%
\begin{align}\label{eq:neq}
 n^{\rm eq}_{\chi}(T)
 %&=\frac{j_{\chi}}{2\pi^2}\int_{m_{\chi}}^{\infty}\sqrt{E_{\chi}^2-m_{\chi}^2}~\text{\rm exp}\left(-\frac{E_{\chi}}{T}\right)E_{\chi}dE_{\chi}\nonumber\\ 
 = \frac{j_{\chi} T^3}{2\pi^2}\left(\frac{m_{\chi}}{T}\right)^2 K_2\l(\frac{m_{\chi}}{T}\r) .
\end{align}
%%%%%%%%%%%%%%%%%%%%%%%%%%%%
Where $m_{\chi}$ and $j_{\chi}$ are the mass and internal degrees of freedom of the dark matter field, respectively. For our entire discussion we assume scalar DM particle with $j_\chi =1$. $T$ is the temperature of the thermal bath during the radiation-dominated era, and $K_2\l({m_{\chi}}/{T}\r)$ is the modified Bessel function of the second kind with order 2. During the RD phase, thermal bath temperature evolves as $T \propto 1/a$, and accordingly the Hubble scale behaves as $H(x)=\Hre\l(\Tre\,x/\bar{a}_{\rm re}\,m_{\chi}\r)^{-2}$. 
Note that apart from conventional thermal bath contribution, we have have an additional non-thermal IR contribution which will be seen to significantly alter the existing DM phenomenology.
With this machinery in hand, we will investigate three possible scenarios of the DM yield interacting with the thermal bath. 
%We will show how the infrared modes gravitationally produced during inflation significantly modifed the existing DM parameter space. 

{\bf Freeze-in Scenario:} In this scenario, the DMs are feebly interacting massive particles (FIMP), and are never in thermal equilibrium with the bath. 
For this case, the Eq.(\ref{eq:nchievolution2}) can therefore be solved analytically assuming two independent contributions from the thermal bath and $Q_{\rm IR}$. 
%In the early time due to large 
The thermal bath contribution comes from,
%energy density DM production rate will be dominant over the IR one.  Therefore, we have
%%%%%%%%%%%%%%%%%%%%%%%%%%%%
\begin{align}\label{eq:nchidynUVFI}
\frac{1}{\bar{a}^3}\frac{d N_\chi}{dt}  = \frac{\langle\sigma v\rangle}{\bar{a}^6}{N_{\chi}^{\rm eq}}^2 .
% & \frac{d}{dx}(Y_{\chi}) \simeq \frac{\langle\sigma v\rangle \,s}{H\,x}{Y^{\rm eq}_{\chi}}^2.
\end{align}
%%%%%%%%%%%%%%%%%%%%%%%%%%%%
In the above equation, we have exploited the condition $Y_{\chi}\ll Y^{\rm eq~ }_{\chi}$ typical to the freeze-in scenario. Second contribution is originated from IR  %by Eq.(\ref{eq:nchievolution3}) is controlled 
rate $Q_{\rm IR}$ as
%%%%%%%%%%%%%%%%%%%%%%%%%%%%%
\begin{align}
\frac{1}{\bar{a}^3}\frac{d N_\chi}{dt}  = 
   Q_{\rm IR}
%&\frac{d}{dx}(Y_{\chi}) \simeq \frac{Q_{\rm IR}(x,\,T)}{H\,s\,x}
\end{align}
%%%%%%%%%%%%%%%%%%%%%%%%%%%%%%%
 and the number density would be same as $N_{\chi}^{\rm IR}$ derived in Eq.(\ref{eq:nchisolIRFI}). 

%that will be dominate over the IR production rate, leading to the UV freeze-in at high temperature. 
%The dominance of the IR production will be seen at late times, when the thermal bath production terminates. The higher the red-tilt, the earlier the dominance of the IR production rate. This leads to the feature of IR freeze-in at late times.\\

\begin{figure}
\begin{center}
\includegraphics[scale=0.255]{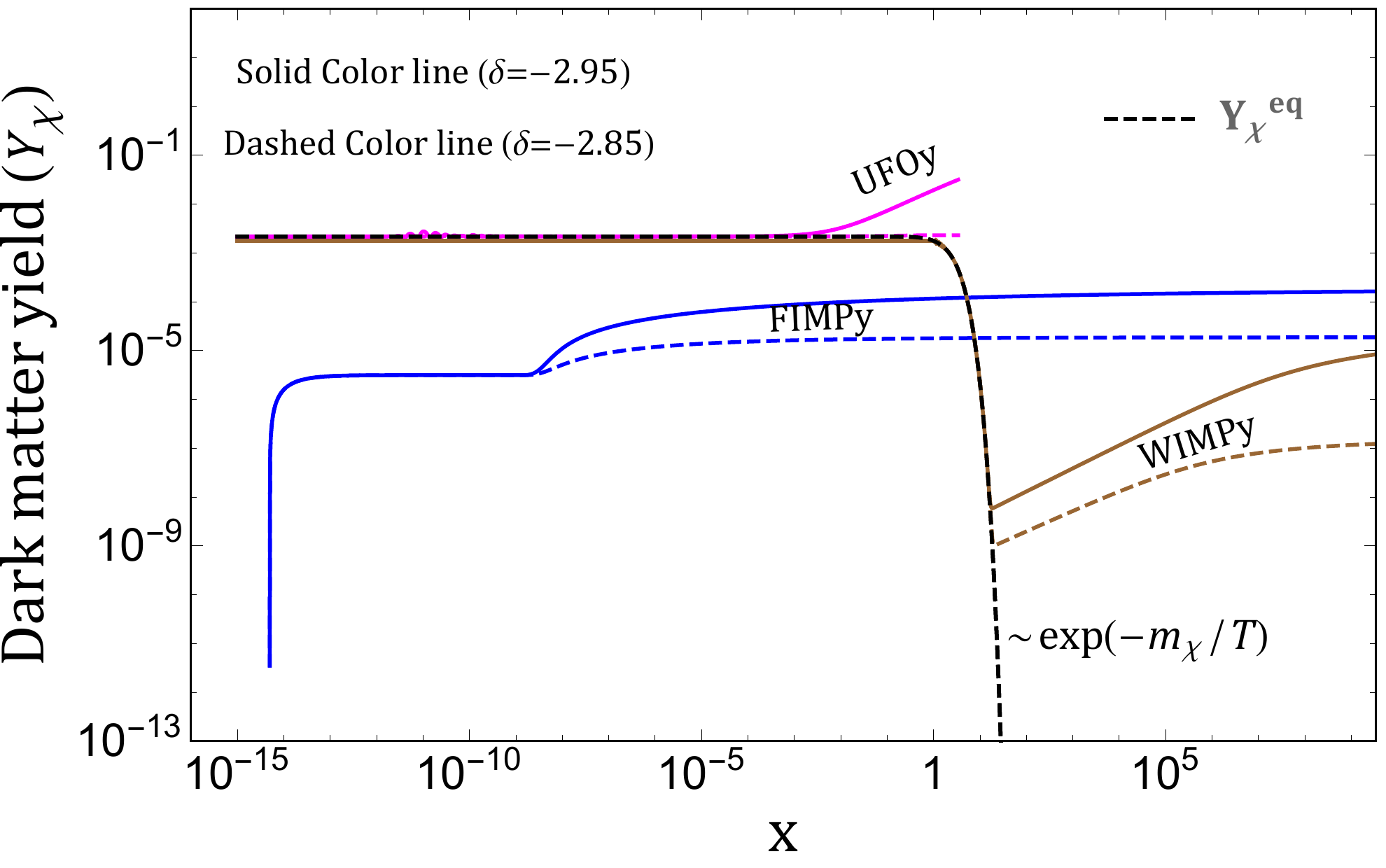}
\caption{\textit{This is the representative figure comparing three different mechanisms of DM yield ($Y_{\chi},~Y_{\chi}^{\rm eq}$) with respect to varying $x={m_{\chi}}/{T}$.
%for different DM mass($m_{\chi}$), spectral index $\delta$, and cross-section $\langle\sigma v\rangle$. 
Solid lines represent the DM yield, $Y_{\chi}$, for UFOy (magenta), FIMPy (blue), and WIMPy (brown) DM, along with the equilibrium evolution $Y_{\chi}^{\rm eq}$ in dashed black curves. 
%and the dashed black lines represent the equilibrium yield $Y_{\chi}^{\rm eq}$. 
%Solid blue line corresponds to FIMP case with the cross-section $\langle\sigma v\rangle= 10^{-35}\,\text{GeV}^{-2}$ and mass $m_\chi = 10$ GeV. Due to the feeble interaction, DM never reaches thermal equilibrium, i.e. $Y_{\chi}< Y_{\chi}^{\rm eq}$. Solid magenta line corresponds to the UFO case with $\langle\sigma v\rangle=  2.56\times 10^{-10}\,\text{GeV}^{-2}$ and mass $m_\chi = 10^{-8}$ GeV. Note that freeze-out happens when $m_{\chi} < T$ is satisfied. Solid brown line corresponds to the WIMP case with $\langle\sigma v\rangle=  5\times 10^{-9}\,\text{GeV}^{-2}$ and mass $m_\chi = 10$ GeV. Unlike the UFO scenario, WIMP freeze-out happens only when $m_{\chi}> T$ is satisfied.    %\textbf{Left panel:} WIMP DM yields for the cross-section $\langle\sigma v\rangle=  5\times 10^{-9}\,\text{GeV}^{-2}$ and mass $m_\chi = 10$ GeV. Note that freeze-out happens only when $m_{\chi}> T$ is satisfied.  
%The mass suppression effect in the number density spectrum of DM field is evident in this figure for both freeze-in and freeze-out scenario.\textbf{Middle panel:} UFOy DM yields for the cross-section $\langle\sigma v\rangle=  2.56\times 10^{-10}\,\text{GeV}^{-2}$ and mass $m_\chi = 10^{-5}$ GeV.for four spectral indices $\delta=(-3, -2.95, -2.9, -2.85)$. \textbf{Right panel:} FIMP DM yields for cross-section $\langle\sigma v\rangle= 10^{-35}\,\text{GeV}^{-2}$, and $m_\chi=10$ GeV. Due to the feeble interaction DM never reaches thermal equilibrium i.e. $Y_{\chi}< Y_{\chi}^{\rm eq}$.
For WIMPy and FIMPy DM, the DM mass is chosen to be $m_{\chi}=10$ GeV, and for UFOy DM, the mass is chosen to be $m_{\chi}= 10^{-8}$ GeV.}}
\label{figcomovingnoFIFO}
\end{center}
\end{figure}

%We call it UV freeze-in 
\underline{FIMPy dark matter:} For this the DM follows the thermal bath evolution, and kinematically stops at the Freeze-in (FI) bath temperature $T_{\rm FI} \simeq m_{\chi}$ with the approximate 
%UV freeze-in 
scale factor $\bar{a}_{\rm FI}\simeq \l(\Tre/m_{\chi}\r)$, considering $\bar{a}_{\rm r e}=1$ for instantaneous transition. During this period, %on the other hand,
%the UV freeze-in process, 
for $m_{\chi}< T_{\rm FI}$, the equilibrium number density (\ref{eq:neq}) is approximate as $n_{\chi}^{\rm eq}\simeq {j_{\chi}T^3}/{\pi^2}$. Utilizing this in Eq.(\ref{eq:nchidynUVFI}), and the boundary condition $N_{\chi}(\bar{a}=\bar{a}_{\rm re} = 1)=0$, we get the FI-DM 
%UV freeze-in 
contribution,
%$N^{\rm FI}_{\chi}= Y^{\rm FI}_{\chi} s(a_{\rm FI}) a_{\rm FI}^3$ as,
%%%%%%%%%%%%%%%%%%%%%%%%%%
\begin{align}\label{eq:nchisolUVFI}
   N^{\rm FI}_{\chi}&=\int_{1}^{\bar{a}_{\rm FI}}\frac{j_{\chi}^2\,\langle\sigma v\rangle\,\Tre^6}{\pi^4\,\He}\l(\frac{d\bar{a}}{\bar{a}^2}\r)\approx \frac{j_{\chi}^2\,\langle\sigma v\rangle\,\Tre^6}{\pi^4\,\He}.
\end{align}
Adding the IR contribution Eq.(\ref{eq:nchisolIRFI}), we, therefore, have the total comoving FIMPy DM particle number density, 
\begin{equation}
    N^{\rm FIMPy}_{\chi} = \l(N^{\rm FI}_{\chi}+N^{\rm IR}_{\chi}\r),
\end{equation}
%%%%%%%%%%%%%%%%%%%%%%%%
constituting early thermal bath FIMP component and late inflationary IR growing component, clearly depicted by the blue line in Fig.(\ref{figcomovingnoFIFO}) for two different sample $\delta = (-2.9,-2.85)$ values.
%\textbf{Freeze out temperature} :
%The freeze-out temperature $T_f$ in general can be computed from Eq.\ref{fo1} by assuming $H(T_{\rm FO}) \propto T_{\rm FO}^2$ as,
%%%%%%%%%%%%%%%%%%%%%%%%%%
%\begin{equation}
%    T_f^{3/2}e^{-m_{\chi}/T_f}=\mathcal{K}(T_{re},T_c)  T_f^k .
%\end{equation}
%%%%%%%%%%%%%%%%%%%%%%%%
%The general solution of the above equation is expressed in terms of Lambert function $W_{-1}(q_1)$ of branch $-1$ with argument $q_1$,
%\begin{equation}{\label{Tf}}
%T_f=-2\,m_\chi\frac{1}{\mathcal{W}_{-1}(q_1)} ~~\mbox{with}~~q_1=-\frac{16\,\pi^3\,\epsilon}{3 M^2_p\langle\sigma v\rangle^2\,j^2_\chi\,m_\chi^2}\,.
%\end{equation}
%%%%%%%%%%%%%%%%%%%%%%%%%%%%%%
%Substituting this expression of $N_{\chi}$, a combination of the UV (\ref{eq:nchisolUVFI}) and IR solutions (\ref{eq:nchisolIRFI}), to the relation (\ref{eq:DMabundance}), and solving the non-linear equation, we find out the relation between the DM mass and interaction cross-section satisfying the current relic.
%Combining Equations (\ref{eq:nchisolUVFI}) and (\ref{eq:nchisolIRFI}),
The present-day relic abundance is computed as
%%%%%%%%%%%%%%%%%%%%%%%%%%%
\begin{align}\label{eq:relicFI}
\Omega^{\rm FIMPy}_\chi h^2= \frac {m_{\chi} N_{\chi}^{\rm FIMPy}}{\epsilon\,\Tre^3\,T_0} \Omega_{\rm R}h^2 .
%&\Omega_\chi h^2=\nno\\
%&\left(\frac{\langle\sigma v\rangle\Tre^3}{\pi^2H_{\rm e}}+ \textcolor{black}{\frac{\mathcal{A}\HI^{7/2}}{4\,\Tre^3\sqrt{m_\chi}}\left(\frac{4m_\chi}{\pi\HI}\right)^{\frac{3+\delta}{2}}\Gamma\left[\frac{3+\delta}{2},\,\frac{\pi\He}{4m_{\chi}\bar{a}_{\rm IR}^2}\right]}\right)\nno\\
%&~\times\left(\frac{m_\chi}{1.94\times10^{-8}\,\mbox{GeV}}\right)\,,
\end{align}
%%%%%%%%%%%%%%%%%%%%%%%%%%
Time evolution of the DM yield conventionally defined as 
$Y_{\chi}\equiv (n_{\chi}/s)$ with the SM entropy density $s(T)=\frac{2\pi^2}{45}g_{\ast s}T^3$, and $g_{\ast s}(T)$ being the number of relativistic entropic degrees of freedom. 
%With this FIMPy nature is depicted in Fig.(\ref{figcomovingnoFIFO}). 
%Hence FIMPy nature of the DM requires even lower thermal bath cross-section to satisfy the abundance except along the standard FIMP slanted black dashed line in the lower part the figure. Beyond this line DM becomes over-abundant shown in gray shaded region.  
For each $\delta$, there exists, therefore, a characteristic DM mass set by the Inert DM mass expressed in Eq.(\ref{eq:mIR}) above which DM becomes overabundant, clearly represented by vertical dotted lines in Fig.
(\ref{fig:DMFIFOfullparameterspace}).

{\bf Freeze-out Scenario:} 
%Another widely accepted dark matter production mechanism in the early universe is the freeze-out mechanism. The key assumption of the freeze-out scenario is that the DM particles will be weakly interacting with the SM particles in the thermal bath. 
%Unlike freeze-in, for this case the interaction strength with the bath would be strong enough for DM being in thermal equilibrium until the background expansion dominates. 
%The produced DM particles in the freeze-out scenario are popularly known as \textit{weakly interacting massive particles}(WIMP). 
Unlike freeze-in, because of the relatively stronger interaction with the thermal bath, the weakly interacting massive DM particles (WIMP) follow the equilibrium number density (\ref{eq:neq}), and well before the thermal freeze-out, it satisfies,
%%%%%%%%%%%%%%%%%%%%%%%%%%%
\begin{equation} \label{eqcond}
2 H = 2\Hre  T^{2} \leq 
\langle\sigma v\rangle n_{\chi}^{\rm eq}(T) % \simeq  \langle\sigma v\rangle \frac{j_\chi \zeta(3)}{\pi^2} T^3~.
\end{equation}
%%%%%%%%%%%%%%%%%%%%%
in the limit $T > m_\chi$. DM particle maintains this condition
until the expansion rate dominates over the interaction rate, which equates $\simeq \langle\sigma v\rangle n_{\chi} \simeq 2H$ at freeze-out temperature $T_{\rm FO}$. 
\textit{Note that IR modes that enter before freeze-out will act as an additional source term in the Boltzmann equation and tend to thermalize, thereby elongating the conventional freeze-out process. We, therefore, have two distinct components to the total DM abundance, constituting the conventional thermal component arising from the freeze-out process and the non-thermal IR component entering the horizon after freeze-out, resulting in the post-freeze-out growth of DM yield(see the Fig.(\ref{figcomovingnoFIFO})), which appears to be a strikingly different feature of the freeze-out process in the present system.} 

Depending on the temperature dependent fall off of the right hand side of the Eq.(\ref{eqcond}), the freeze-out condition, $ \langle\sigma v\rangle n_{\chi}^{\rm eq}(T_{\rm FO}) = 2 H(T_{\rm FO})$ can be satisfied for both non-relativistic $m_{\chi} > T_{\rm FO}$ (WIMPy), and relativistic $m_{\chi} < T_{\rm FO}$ (UFOy) DM mass range. For simple phenomenological purposes, we assume $\langle\sigma v\rangle$ is temperature independent.
%%%%%%%%%%%%%%%%%%%%%%%%%%%%%%%%
\underline{WIMPy dark matter:}
In these scenario the DM freezes-out from the thermal bath non-relativistically. 
Boltzmann suppression becomes predominant in the equilibrium number density expression (\ref{eq:neq}). Hence, the freeze-out temperature, we call it $T_{\rm WFO}$ can be approximately calculated from
%%%%%%%%%%%%%%%%%%%%%%%%%%%%%%
\begin{equation}
 2\Hre  T_{\rm WFO}^{2} \simeq  \langle\sigma v\rangle \frac {j_\chi}{8\pi^{\frac32}} (2m_\chi T_{\rm WFO})^{\frac32} e^{-\frac{m_\chi}{T_{\rm WFO}}}.
\end{equation}
%%%%%%%%%%%%%%%%%%%%%%%%%%%%%%
After freezing out, the WIMPy part of the DM is then governed by
%%%%%%%%%%%%%%%%%%%%%%%%%%%
\begin{equation} \label{WIMPlate}
\frac{1}{\bar{a}^3}\frac{d N_\chi}{dt}  \simeq \underbrace{-\frac{\langle\sigma v\rangle}{\bar{a}^6} N_{\chi}^2}_{\text{ annihilation term}}
  + Q_{\rm IR}.
   %\frac{d N_{\chi}}{dx} \simeq  -\frac{\lambda}{x^2} N_{\chi}^2.
\end{equation}
%%%%%%%%%%%%%%%%%%%%%%%%%%%
%Where, $\lambda =\frac{8 \MP^2 g_{\ast s} \langle\sigma v\rangle}{15 g_{\ast} m_\chi} $, and $x\equiv \l(m_{\chi}/T\r)$. 
Without IR source term, above equation yields the well known freeze-out components $N^{\rm WFO}_{\chi}= Y^{\rm WFO}_{\chi} s(a_{\rm 0}) a_{\rm 0}^3 \propto  \frac{1}{\langle\sigma v\rangle} \frac{m_\chi}{T_{\rm WFO}}$. However due to the continuous IR injection, the  annihilation term remains active even after the freeze-out for the present case, and their interplay eventually leads the modified final WIMPy DM yield (see Fig.(\ref{figcomovingnoFIFO})) consisting of
%Adding the non-thermal IR contribution Eq.(\ref{eq:nchisolIRFI}), 
%we, therefore, have the total comoving WFOy DM 
both the components.%$N^{\rm WFOy}_{\chi} = \l(N^{\rm WFO}_{\chi}+N^{\rm IR}_{\chi}\r)$, 
%giving rise to the present-day relic abundance as
%%%%%%%%%%%%%%%%%%%%%%%%%%%
\begin{equation}\label{eq:relicFO}
\Omega^{\rm WIMPy}_\chi h^2= \frac {m_{\chi} N_{\chi}^{\rm WFOy}}{\epsilon\,\Tre^3\,T_0} \Omega_{\rm R}h^2 
\end{equation}
%%%%%%%%%%%%%%%%%%
The time evolution of the DM yield with this WIMpy nature is clearly depicted in Fig.(\ref{figcomovingnoFIFO}) as a brown solid line for two sample values of $\delta = (-2.95,-2.85)$. It shows how the IR component contributes notably after freeze-out.

\underline{UFOy dark matter:} 
Within this framework, the ultra-relativistic freeze-out(UFO) appears as a prominent intermediate phase, which causes the seamless WIMP to FIMP transition in the low DM mass regime(see Fig.(\ref{fig:DMFIFOfullparameterspace})). Unlike the WIMP case, UFO happens much before the Boltzmann suppression becomes dominant in the equilibrium number density expression (\ref{eq:neq}), Hence, the freeze-out temperature, which we call $T_{\rm UFO}$, can be approximately calculated from Eq.(\ref{eqcond}),
%%%%%%%%%%%%%%%%%%%%%%%%%%%%%%%%
\begin{equation}
  T_{\rm UFO} \simeq \frac{2 \Hre\pi^2}{\langle\sigma v\rangle\Tre^2\, j_\chi\zeta(3)},
\end{equation}
%%%%%%%%%%%%%%%%%%%%%%%%%%%%%
where use has been made of $ n_{\chi}^{\rm eq}(T)\simeq  \langle\sigma v\rangle \frac{j_\chi \zeta(3)}{\pi^2} T^3~$ in the relativistic regime. Such a scenario is first realized in the case of neutrino decoupling. Recently, it has been noted and extensively studied in the DM context in \cite{Henrich:2025gsd, Henrich:2025sli}. Nevertheless, as the freeze-out happens when the DM particles are still in the relativistic regime, one can immediately compute the approximate UFO contribution to the comoving DM number as,
%%%%%%%%%%%%%%%%%%%%%%%%%%
\begin{equation}
    N^{\rm UFO}_{\chi} \simeq a^3 n^{\rm eq}_{\chi}({T_{\rm UFO}}) = \frac {j_\chi\zeta(3)}{\pi^2} .  
\end{equation}
%%%%%%%%%%%%%%%%%%%%%%%%%%%%%%
%\textcolor{magenta}{\textcolor{blue}{Suggest modifications}This is not correct, due to strong interaction IR production also affected; numerically, we have checked that the simple sum is not correct.}
Due to relativistic freeze-out, therefore, UFO evolution of the DM yield cannot be approximated as the WIMPy case. We, therefore, have solved  Eq.(\ref{eq:nchievolution2}) numerically without any approximation 
%Adding the IR contribution Eq.(\ref{eq:nchisolIRFI}), we, therefore, have the total comoving UFOy DM particle number density, $N^{\rm UFOy}_{\chi} \simeq \l(N^{\rm UFO}_{\chi}+N^{\rm IR}_{\chi}\r)$, 
for the present-day relic abundance
%%%%%%%%%%%%%%%%%%%%%%%%%%%
\begin{equation}\label{eq:relicUFO}
\Omega^{\rm UFOy}_\chi h^2= \frac {m_{\chi} N_{\chi}^{\rm UFOy}}{\epsilon\,\Tre^3\,T_0} \Omega_{\rm R}h^2\,. 
\end{equation}
%%%%%%%%%%%%%%%%%%%%%%%%%%%%%%%
Time evolution of the DM yield with this UFOy nature is nicely depicted in Fig.(\ref{figcomovingnoFIFO}) in magenta lines for two different sample $\delta = (-2.95,-2.85)$ values.
Post freeze-out yield surpassing $n_{\chi}^{\rm eq}$ emerges as an interesting characteristic feature of UFO DM, as opposed to the picture presented in \cite{Henrich:2025gsd, Henrich:2025sli}, where post-freeze-out yield never exceeds $n_{\chi}^{\rm eq}$ due to the continuous injection of bath particles during finite reheating.
%%%%%%%%%%%%%%%%%%%%%%%%%%%%%%%%%%

%The expression of DM relic abundance in terms of radiation abundance is given by 
%%%%%%%%%%%%%%%%%%%%%%%%%%\begin{equation}\label{eq:DMabundance}\Omega_{\chi}h^2= \frac{  m_{\chi}N_{\chi}(\bar{a}_{\rm F})}{\epsilon\, T_{\rm F}^3\,\bar{a}^3_{\rm F}\,T_0}\Omega_{\rm R}h^2=0.12\,,\end{equation}
%%%%%%%%%%%%%%%%%%%%%%%%%%%%%where $T_{\rm F}$ is the temperature of the radiation bath at very late time $\bar{a}_{\rm F}$ during the RD phase when both radiation and DM energy density freeze, the dimensionless present-day radiation energy density, $\Omega_{\rm R}h^2=4.3\times 10^{-5}$ \cite{sdf}, and the present-day CMB temperature is $T_0\simeq2.35\times 10^{-13}$ GeV.\\ However, to make the DM mass $m_{\chi}$ vs $\langle\sigma v\rangle$ parameter space by obeying the current relic in the freeze-out case, we work with the DM yield. To get the present DM relic abundance $\Omega_{\chi}h^2\simeq 0.12$ \cite{Planck:2018vyg}, we require \cite{Bernal_2022}
%%%%%%%%%%%%%%%%%%%%%%%%%%%\begin{equation}\label{eq:mYF}m_{\chi}\,Y_{\chi}(A_{\rm F})=\frac{1}{s_0}\frac{\rho_c}{h^2}\Omega_{\chi}h^2\simeq 4.68\times 10^{-10} \text{GeV}\,,\end{equation}
%%%%%%%%%%%%%%%%%%%%%%%%%%%%%where $\rho_c\simeq 1.05\times 10^{-5}\,h^2$ GeV/$\text{cm}^3$ being the critical energy density, and $s_0\simeq 2.69\times 10^3\,\text{cm}^{-3}$ is the present entropy density \cite{ParticleDataGroup:2024cfk}.\\ 
%%%%%%%%%%%%%%%%%%%%%%
%%%%%%%%%%%%%%
\subsection{Results and discussions}

\begin{figure}
\begin{center}
\includegraphics[scale=0.40]
{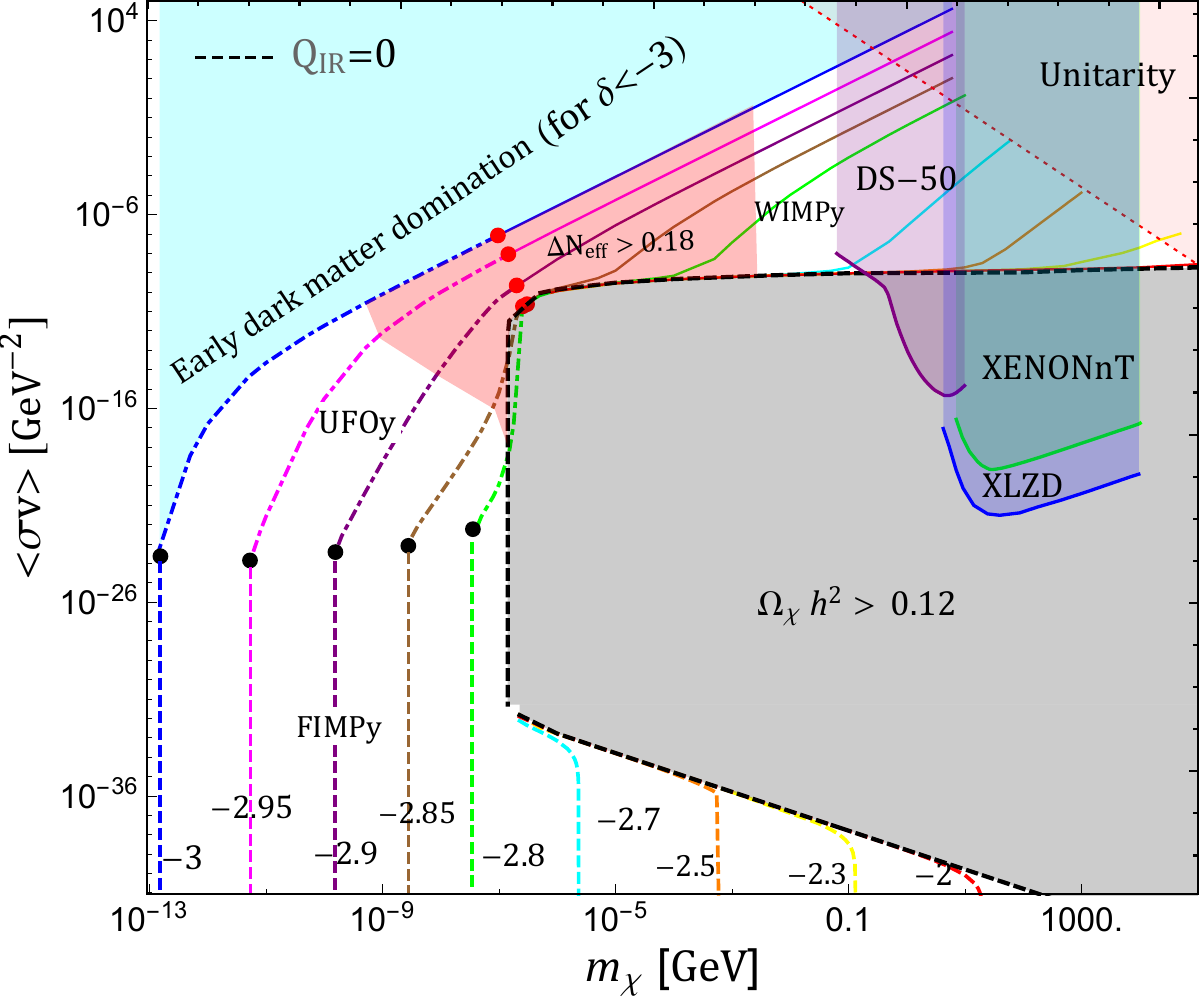}
%{figDMfullparameterspace.pdf}
\caption{\textit{Figure represents the ($m_{\chi},\,\langle\sigma v\rangle$) DM parameter space for different $\delta$ values exploring three distinct possibilities (WIMPy, FIMPy, UFOy) of the DM production mechanism. Five  \textcolor{red}{red} dots on the solid colored lines indicate the transition from WIMPy  (solid colored lines) to UFOy (dot-dashed colored lines) scenarios, and five \textbf{black} dots indicate the transition from the UFOy to FIMPy (dashed colored lines) scenarios. For $\delta=-3, -2.95, -2.9, -2.85, -2.8$, $Q_{\rm IR}$ gives the total abundance of DM in the non-thermal freeze-in scenario, resulting in the sharp fall of $\langle\sigma v\rangle$ from the black dots at those critical mass values $m_{\chi}^{\rm IR}$(see Eq.(\ref{eq:mIR})). The light red shaded region} $\l(m_{\chi}\in (m_{\chi}^{\rm UFO},\,4\,{\rm MeV}]\r)$ \textit{is excluded by the $\Delta N_{\rm eff}$ constraint. 
%and Lyman-$\alpha$ bounds, respectively. 
}}
\label{fig:DMFIFOfullparameterspace}
\end{center}
\end{figure}

%\textcolor{red}{Brief discussion on Isocurvature \cite{Chakraborty:2024rgl, Chakraborty:2025oyj}}
Time evolution would be non-trivial for DM when it is interacting with the thermal bath. Fig.(\ref{figcomovingnoFIFO}) clearly represents the DM yield evolution for three distinct cases discussed above. %For both freeze-in and freeze-out scenarios, we numerically solve the dynamical equation (\ref{eq:nchievolution4}) with $Y_{\chi}(x_{\rm int})=0$. In the freeze-in case, for a given mass $m_{\chi}$, we set the initial point of evolution $x_{\rm int}=\l(m_{\chi}/\Tre\r)$, which is a very small number for the sub-Hubble masses. In the freeze-out case, we set $x_{\rm int}=1$, as non-relativistic WIMP-like freeze-out always occurs when $x>>1$.
%For our present purpose, we assume real scalar DM with $j_\chi=1$. 
The Fig.(\ref{figmvsdeltapuregrav}) represents the parameter space for inert DM (without any thermal bath interaction), where the DM mass is completely fixed for a given spectral index parameter $\delta$. On the other hand when DM interacting with the thermal bath, the phenomenology becomes extremely rich as depicted in Fig.(\ref{fig:DMFIFOfullparameterspace}). The figure in different colored lines captures the complete parameter space for DM yield clearly distinguishing three types namely, WIMPy (solid) , FIMPy (dashed), and continuously connected by UFOy (dot-dashed) dark matter. We assumed instantaneous reheating case with $T_{\rm re} \simeq 10^{15}$ GeV. Without any IR contribution ($Q_{\rm IR}=0$) black dashed line represents the well known parameter space for FIMP (Lower slanted part) and WIMP miracle (upper horizontal part) smoothly joined by vertical line \cite{Haque:2023yra}. However, we emphasize the fact that the inflationary gravitational production is inescapable and it is the inflationary IR production $Q_{\rm IR}$ which may largely dominate the DM yield. Indeed all the colored lines represents the fact that the gravitational IR contribution ($Q_{\rm IR} \neq 0$) to the DM yield is mostly the dominant one, and they come with different species of DM arising depending on their mass and cross-section. As one can realize, parameter space is greatly influenced by IR spectral tilt $\delta$. Higher the tilt, larger would be the IR contribution to the total DM yield, and giving rise to maximum deviation from the standard FIMP/WIMP paradigm. %which requires larger $\langle\sigma v\rangle$ value for a given DM mass to set correct relic abundance.
%non-zero $Q_{\rm IR}$, each solid colored line grows out of the horizontal pure radiation line, and with the increase of DM mass, it requires a larger interaction strength to achieve a prolonged thermal equilibrium for the DM species to produce the correct relic in the presence of a significant IR production rate. The higher the IR tilt, the stronger the $Q_{\rm IR}$, the larger the $\langle\sigma v\rangle$ for a given DM mass. 
For example, in the standard scenario without IR contribution, the black dashed lines are the only allowed parameter, and the minimum FIMP/WIMP DM mass possible would be of the order of $\sim 10^{-7}$ GeV (vertical part of dashed black line). %below which its yield becomes under abundant. 
However, inflationary IR production enables the mass of the Inert/FIMPy DM down to $\sim 10^{-13}$ GeV for $\delta = -3$, which, indeed, opens up a large region of allowed DM parameter space
%Neglecting the IR contribution, the dark matter mass($m_{\chi}$) vs annihilation cross section($\langle\sigma v\rangle$) variation follows the slanted black dashed line in the lower part of Fig.(\ref{fig:DMFIFOfullparameterspace}). 
%\textcolor{red}{The maximum dark matter mass that can account for the current relic abundance is of order $\mathcal{O}(10^4)$ GeV}. 
as nicely depicted the Fig.(\ref{fig:DMFIFOfullparameterspace}) for different $\delta$ values between $[-3,2]$. 

We numerically find a critical $\delta_{\rm c} \simeq -2.758$ such that for $\delta < \delta_c$, upon increasing $\langle \sigma v \rangle$, line smoothly transits  from Inert $\rightarrow$ FIMPy $\rightarrow$ UFOy $\rightarrow$ WIMPy as shown for $\delta = (-3, -2.95, -2.9, -2.85, - 2.8)$. For these cases, it is the IR contribution that dominates the DM abundance for all mass range. On the other hand, for $\delta > \delta_c$, upon increasing $\langle \sigma v \rangle$, line smoothly transits  from Inert $\rightarrow$ FIMPy $\rightarrow$ FIMP $\rightarrow$ WIMP $\rightarrow$ WIMPy as shown for $\delta = (-2.7, -2.5, -2.3)$. For these cases all WIMPy and FIMPy lines merge with the standard black dashed WIMP and FIMP line, and IR contribution to the abundance dominates for higher DM masses only.

\underline{{\bf Lyman-$\alpha$ constraints:}} 
%The low-mass dark matter produced by ultra-relativistic freeze-out may cause serious problem in the formation of small-scale structures if they remain sufficiently warm at the time of structure formation, which happens around the temperature $T\simeq 1$ eV. 
The very low-mass DM produced in the early universe is subjected to tight constraints by Lyman-$\alpha$ observation. For example, UFO-type DM produced during standard radiation domination %encounters a serious observational challenge owing to its warmness, which is incompatible with the small-scale structure formation around $T\simeq 1$ eV. Hence, 
are shown \cite{ Henrich:2025gsd, Henrich:2025sli} to be in conflict with such observation.
%the large-scale structure formation constraints as well as the Lyman-$\alpha$ data
%and potentially rule out this possibility of UFO in the standard scenario,
%as pointed out in . 
However, our present UFOy DM has significant non-thermal IR components, originating from inflationary gravitational production.
%there is yet a glimpse of hope for the possibility of UFOs during the radiation-dominated era. 
%The late-time growth of $Y_{\chi}$, due to the significant IR production rate, perhaps saves this possibility from being completely excluded by observations. In the present scenario, 
%the intermediate UFO phase appears solely due to the substantial IR contribution $N_{\chi}^{\rm IR}$ to the DM number density. 
%after ultra-relativistic freeze-out at an instant, say, $a_{\rm UFO}$, the continuous growth of $Y_{\chi}$ 
%caused by the gradual entry of the low-energy modes up to $a_{\rm IR}$. 
%Therefore, the saturated total number density $Y_{\chi}$ accounting for the current relic is dominated by the low-energy infrared contribution. 
Assuming a particular DM mode $k_{T}=a(T)H(T)$ entering the horizon at a temperature $T$ provides the highest contribution to $N_{\chi}$, and the velocity of DM  associated with it
%the physical momentum of $k_T$
must satisfy 
the Lyman-$\alpha$ constraint around $T\simeq 1$ eV. We can then write 
%%%%%%%%%%%%%%%%%%%%%%%%%%%%%
\begin{align}\label{eq:Lymanalpha}
&\frac{k_T}{a_{\rm IR}m_{\chi}}%\Rightarrow\Hre\l(\frac{1\,{\rm eV}}{\Tre}\r)\l(\frac{T}{\Tre}\r)\frac{1}{m_{\chi}}< 2\times 10^{-4}\nno\\
=\Hre\l(\frac{1\,{\rm eV}}{\Tre}\r)\l(\frac{1}{x\,\Tre}\r)< 2\times 10^{-4}\,,
%&\Rightarrow x\gtrsim \l( 7\times 10^{-24}\r)\,,
\end{align}
%%%%%%%%%%%%%%%%%%%%%%%%%%%%%%%%%
For instantaneous reheating $\Tre\simeq 10^{15}$ GeV, one obtains $x\gtrsim  7\times 10^{-24}$. We find that the IR contribution remains predominant up to a larger value of $x$, respecting the above relation (\ref{eq:Lymanalpha}) for $-3 < \delta < \delta_c$.
%, as can be seen in the middle panel of Fig.(\ref{figcomovingnoFIFO}). 
For IR tilt ($\delta\simeq -3$), 
assuming dominant contribution from the largest scale reentering the horizon around $T\simeq 1$ eV, the relation (\ref{eq:Lymanalpha}) yields the lowest possible DM mass to be 
%subject to the Lyman-$\alpha$ bound, which turns out to be 
$m_{\chi}\gtrsim (7\times 10^{-33})$ GeV,  and it is indeed well below the critical IR mass $\mathcal{O}(10^{-13})$ GeV we obtained for $\delta=-3$, satisfying the DM abundance. Therefore, all the infrared modes that significantly contribute to the total DM number density will remain cold enough to be compatible with the Lyman-$\alpha$ constraint. This prescription is likewise applicable to the critical IR mass $m_{\chi}^{\rm IR}$ produced in the non-thermal FIMP scenario. It is the significant IR production caused by low-energy mode entry that allows these low DM masses($m_{\chi}\lesssim 5$ keV), produced in the freeze-in and UFO mechanisms during the post-inflationary radiation-dominated universe, as opposed to the standard non-thermal FIMP and thermal UFO scenarios, where DM masses $m_\chi\lesssim5$ keV are ruled out by the Lyman-$\alpha$ bound for instantaneous reheating.

%%%%%%%%%%%%%%%%%%%%%%%%%%%%%%
%%%%%%%%%%%%%%%%%%%%%%%%%%%%%%%%%
%%%%%%%%%%%%%%%%%%%%%%%%%%%%
%corresponding to $\delta= -3, -2.95, -2.9 -2.85, -2.8$
%\subsection{$\Delta N_{\rm eff}$ constraint:}
\underline{\textbf{$\Delta N_{\rm eff}$ constraints:}}
At the time of Big Bang nucleosynthesis(BBN), it is essential to ensure that any contribution from BSM physics to the total radiation energy density must satisfy the observational upper bound on the effective number of relativistic degrees of freedom, conventionally parametrized as
%Such additional relativistic degrees of freedom are conventionally expressed in terms of the effective number of neutrino species, $ N_{\text{eff}}$. The deviation from the SM prediction (which corresponds to three active neutrino species) is defined as 
$\Delta N_{\text{eff}} = %N_{\text{eff}}-3.044=
(43/7)\,\rho_\chi/\rho_{\rm rad} < 0.18 $ \cite{Jinno_2012, Haque:2023yra, Bennett:2020zkv, Froustey:2020mcq,Akita:2020szl,Chakraborty:2023ocr} 
%The current observational constraint is $
%\Delta N_{\text{eff}} < 0.18$ 
at $95\%$ confidence level (CL)\cite{ACT:2025tim, ACT:2025fju}, where $\rho_{\chi}$ and $\rho_{\rm rad}$ are energy densities of DM and radiation background, respectively. In case of UFO, the DM masses for which the condition $T_{\rm UFO}\gg4\,{\rm MeV}\gg m_{\chi}$ is satisfied, may contribute to the $\Delta N_{\text{eff}}$ constraint at the time of BBN, owing to the continuous energy injection in the radiation bath by $Q_{\rm IR}$ from $T_{\rm UFO}$ to $T_{\rm BBN}\simeq 4$ MeV. Using the IR spectrum (\ref{exactbeta}), this constraint($\Delta N^{\rm BBN}_{\text{eff}}$) at the time of BBN can be expressed as
%The UFO like FO, if the DM freezes out before the BBN, its temperature at the BBN,
%%%%%%%%%%%%%%%%%%%%%%%%%%%%%
\begin{align}\label{eq:deltaneff}
\Delta N^{\rm BBN}_{\text{eff}}&=\l(\frac{43}{7}\r)\frac{\int^{k_{\rm UFO}}_{k_{\rm BBN}}k^2 m_{\chi}|\beta_k|^2\,dk}{6\pi^2\MP^2 a_{\rm BBN}^3H^2(T_{\rm BBN})}~<0.18\,,\nno\\
&\simeq\l(\frac{43\Hre^2}{42\pi^2\MP^2}\r)\l(\frac{g_{\ast}(\Tre)}{g_{\ast}(T_{\rm BBN})}\r)^{1/3}\l(\frac{\Tre}{T_{\rm BBN}}\r)\sqrt{\frac{m_{\chi}}{\Hre}}\nno\\
&\times
\begin{cases}
{\rm  ln}\l(\frac{T_{\rm UFO}}{4\,{\rm MeV}}\r)~~~~\mbox{for}~~\delta=-3\\
\frac{\l(\frac{T_{\rm UFO}}{10^{15}\,{\rm GeV}}\r)^{\delta +3}}{(\delta+3)}~~\mbox{for}~~|\delta|<3\,  
\end{cases}
< 0.18\,,
  %T_\chi^{\rm BBN}=\frac{a_{\rm UFO}}{a_{\rm BBN}}\,T_{\rm UFO}=
\end{align}
%%%%%%%%%%%%%%%%%%%%%%%%%%%%%%%
where we take $g_{\ast}(\Tre)=106.75$ and $g_{\ast}(T_{\rm BBN})=10.75$. It is found that for $|\delta|\lesssim 3$, the constraint (\ref{eq:deltaneff}) is respected in the mass range $[m^{\rm IR}_{\chi},\,m^{\rm UFO}_{\chi}]$, where $m^{\rm UFO}_{\chi}$ lies in the range $[10^{-10},\,10^{-7}]$ GeV for $\delta\in [-3,\,-2.8]$. For any $|\delta|\lesssim 3$, the scalar DM masses in the range $m_{\chi}\in (m_{\chi}^{\rm UFO},\,4\,{\rm MeV}]$ are incompatible with the current $\Delta N_{\text{eff}}$ constraint both in WIMP and UFO scenarios, as they remain in equilibrium with the thermal bath at $T_{\rm BBN}$, always behaving like an additional degree of freedom\cite{Haque:2023yra}. For instance, $\delta=-2.9$, in the UFO mechanism, the allowed mass range is found to be $[m_{\chi}^{\rm IR},\,10^{-8}]$ GeV, and in the WIMP mechanism, it yields $m_{\chi}\gtrsim 4$ MeV. Therefore, subject to Lyman-$\alpha$ and $\Delta N_{\text{eff}}$ constraints, the admissible relativistic DM mass range is found to be $[m^{\rm IR}_{\chi},\,m^{\rm UFO}_{\chi}]$, and the allowed non-relativistic mass range is $m_{\chi}\geq 4$ MeV in UFO and WIMP pictures. Furthermore, we have checked that for FIMP, no DM parameter space is excluded by the $\Delta N_{\rm eff}$ constraints. Therefore, subject to Lyman-$\alpha$ and $\Delta N_{\text{eff}}$ constraints, the FIMP having masses $m_\chi>5$ keV are allowed.   

%\iffalse

\section{Acknowledgments}
AC gratefully acknowledges the Ministry of Human Resource Development, Government of India (GoI), for financial assistance. RM acknowledges the support of
postdoctoral research fellowship (PDRF) from IISER Berhampur. DM acknowledges support by Institut Pascal at Université Paris-Saclay for attending 
the Paris-Saclay Astroparticle Symposium 2025. DM also thanks Prof. Yann Mambrini and Dr. Riajul Haque for fruitful discussions during the Astroparticle Symposium.
%DM wishes to acknowledge support from the Science and Engineering Research Board~(SERB), Department of Science and Technology~(DST), Government of India~(GoI), through the Core Research Grant CRG/2020/003664.

\bibliographystyle{apsrev4-1}
\bibliography{AYANreferences}

\end{document}